\documentclass[twocolumn,prl,aps]{revtex4-1}
\usepackage[utf8]{inputenc}
\usepackage{graphicx}
\usepackage{hyperref}
\usepackage{amssymb}
\usepackage{amsmath}
\usepackage{xcolor}
\usepackage{lineno}
\usepackage{enumitem}
\usepackage{endnotes}

\usepackage{booktabs}

\newcommand{\appropto}{\mathrel{\vcenter{
  \offinterlineskip\halign{\hfil$##$\cr
    \propto\cr\noalign{\kern2pt}\sim\cr\noalign{\kern-2pt}}}}}

\begin{document}

\title{Determining the fraction of cosmic-ray protons at ultra-high energies with cosmogenic neutrinos} 

\author{Arjen van Vliet$^{1,2}$}
\thanks{arjen.van.vliet@desy.de}
\author{Rafael \surname{Alves~Batista}$^3$}
\author{J\"org R. H\"orandel$^{1,4,5}$}
\affiliation{
$^1$ Radboud University, Department of Astrophysics/IMAPP, P.O. Box 9010, 6500 GL Nijmegen, The Netherlands \\
$^2$ Deutsches Elektronen-Synchrotron (DESY), Platanenallee 6, 15738 Zeuthen, Germany \\
$^3$ Universidade de São Paulo, Instituto de Astronomia, Geofísica e Ciências Atmosféricas, Rua do Matão, 1226, 05508-090, São Paulo-SP, Brazil \\
$^4$ NIKHEF, Science Park Amsterdam, 1098 XG Amsterdam, The Netherlands\\
$^5$ Vrije Universiteit Brussel, Dept. of Physics and Astronomy, B-1050 Brussels, Belgium
}

\begin{abstract}

Cosmogenic neutrinos are produced when ultra-high-energy cosmic rays (UHECRs) interact with cosmological photon fields. Limits on the diffuse flux of these neutrinos can be used to constrain the fraction of protons arriving at Earth with energies $E_{p} \gtrsim  30 \; \text{EeV}$, thereby providing bounds on the composition of UHECRs without fully relying on hadronic interaction models. We show to which extent current neutrino telescopes already constrain this fraction of protons and discuss the prospects for next-generation detectors to further constrain it. Additionally, we discuss the implications of these limits for several popular candidates for UHECR source classes. 

\end{abstract}



\maketitle

Understanding the composition of ultra-high-energy cosmic rays (UHECRs) is crucial to elucidate their origin, which remains an open problem in astrophysics. This is especially true at the highest energies ($E \gtrsim 30 \; \text{EeV}$) as there protons are expected to point back to their sources, while heavier nuclei might still be deflected significantly by Galactic and extragalactic magnetic fields (see e.g. Refs.~\cite{Farrar:2017lhm,AlvesBatista:2017vob}). However, due to the small flux of cosmic rays at these energies and the dependence on hadronic interaction models, it is challenging to determine the composition reliably. The Pierre Auger Observatory and the Telescope Array (TA) have nonetheless been able to provide first indications of what the composition at these energies might be~\cite{Aab:2014kda,Aab:2014aea,ObservatoryMichaelUngerforthePierreAuger:2017fhr,Abbasi:2018nun}. Efforts to improve on this are underway with planned upgrades of both Auger (AugerPrime)~\cite{Aab:2016vlz} and TA (TAx4)~\cite{Kido:2017nhz}.

The measurements by Auger indicate that the depth of the shower maximum ($X_\text{max}$) -- a proxy for the composition --  favors the interpretation of a very light composition at $\sim 2\;\text{EeV}$, dominated by protons, with the average mass composition increasing with energy after that. This increase seems to stop, however, at an energy of  $\sim 50\;\text{EeV}$~\cite{Aab:2017cgk}, which might be an indication for a subdominant light component at these energies. TA's measurements of $X_\text{max}$ are compatible with the results found by Auger within uncertainties~\cite{deSouza:2017wgx}. However, as TA has collected a smaller number of cosmic rays and is also less sensitive to the composition than Auger, a larger range of composition scenarios (even a pure-proton case~\cite{Abbasi:2018nun}) is still possible in the northern hemisphere. 

Additionally, predictions by different air-shower simulation models leave room for varying interpretations of the data. State-of-the-art hadronic-interaction models such as Sibyll2.3c~\cite{Riehn:2017mfm}, EPOS-LHC~\cite{eposlhc} and QGSJetII-04~\cite{qgsjetii} have been designed taking into account LHC data. However, the required extrapolation in energy and phase space to simulate UHECR air showers leaves room for disagreement. The current problems include a significantly larger hadronic component, which manifests itself as a surplus of muons at ground compared to the simulations~\cite{Aab:2014dua,Aab:2014pza,Aab:2016hkv}. An interaction-model independent probe of the composition is, therefore, very desirable.

We present here a new method to constrain the composition, in particular the fraction of protons at Earth ($f$), at $E \gtrsim 30 \; \text{EeV}$ without relying on air-shower observables. This method is based on the (non-)observation of cosmogenic neutrinos. 

Cosmic-ray interactions with the cosmic microwave background (CMB) and the extragalactic background light (EBL) can produce neutrinos of cosmogenic origin. Additionally, unstable atomic nuclei, produced during photodisintegration or photo-pion production, may undergo nuclear decay and produce cosmogenic neutrinos. 

Recently, much effort has been put into interpreting the spectrum and composition measurements in terms of simple astrophysical models. To this end, the main ingredients needed relate to source properties: spectral index, maximal energy attainable, abundance of each nuclear species, luminosity, distribution, and evolution. The combined spectrum-composition fits performed in Refs.~\cite{Taylor:2015rla,Aab:2016zth,Romero-Wolf:2017xqe,AlvesBatista:2018zui} are first approaches to the problem and use a number of simplifying assumptions. These fits, in most cases, find as best fit an intermediate to heavy composition (dominated by nuclei with charge $Z \geq 6$), a relatively hard injection spectrum (spectral index $\alpha \lesssim 1.3$), and a relatively low maximum rigidity ($R_\text{max}\equiv E_\text{max} / Z \lesssim 7$ EV, where $E_\text{max}$ is the maximum energy) of the sources. 
Under these assumptions, the expected cosmogenic neutrino flux is so low that it will be hard to detect even for planned neutrino detectors as ARA~\cite{ara2012a}, ARIANNA~\cite{arianna2015a} and GRAND~\cite{Alvarez-Muniz:2018bhp} (see Refs.~\cite{vanVliet:2017obm,Romero-Wolf:2017xqe,AlvesBatista:2018zui,heinze2018}). 

However, these combined-fit studies assume a continuous distribution of identical sources and rigidity-dependent maximum energies. Under these assumptions, with a mass composition that is getting increasingly heavier with energy, it is not possible to obtain even a subdominant proton contribution at the highest energies. Just from the measured $X_\text{max}$ data, however, such a proton component could be present. If, for example, one would drop the assumption that all sources in the Universe are identical,  such protons could be produced by a source that can accelerate cosmic rays up to extremely high energies, but only gives a subdominant contribution to the full UHECR spectrum.

While such a subdominant proton component has a limited effect on the spectrum and composition, it strongly alters the expected cosmogenic neutrino flux. This is because protons produce significantly more neutrinos when propagating through the Universe than heavier nuclei (see e.g. Refs.~\cite{Kotera:2010yn,Roulet:2012rv}). 

Note that in Ref.~\cite{Eichmann:2017iyr} combined fits to the UHECR spectrum and composition are presented for a detailed model of radio galaxies that does not assume identical sources or a continuous source distribution. The spectrum, composition and large-scale arrival distribution of UHECRs are well reproduced in this work for sources with a predominantly light composition and a spectral index of $\alpha \approx 1.8$. 

We employ the CRPropa 3 code~\cite{Batista:2016yrx} to simulate the propagation of UHE protons and secondary neutrinos. This includes all relevant interactions and energy-loss processes, namely: photo-pion production, pair production, and nuclear decay, as well as adiabatic losses due to the expansion of the Universe. 

Our standard scenario consists of a distribution of homogeneous and identical sources with the same luminosity, extending up to a redshift of $z_\text{max} = 4.0$. The source evolution (SE) is a combination of the evolutions of both the source number density and luminosity, and is given by:
\begin{equation}
\text{SE}(z) = 
	\begin{cases}
	  (1 + z)^m & \text{for } m \leq 0 \\
	  (1+z)^m & \text{for } m > 0 \text{ and }  z < 1.5  \\
      2.5^m & \text{for } m > 0 \text{ and } z \geq 1.5
	\end{cases},
  \label{eq:sourceevolution0}
\end{equation}
where $m$ is the source-evolution parameter. In reality, the evolution of most source candidates is complex and cannot be expressed with a single parameter. For typical source candidates the evolution grows up to a given redshift $1.0 \lesssim z_1 \lesssim 1.7$, reaches a plateau (or increases very slowly) between $z_1$ and $2.7 \lesssim z_2 \lesssim 4.0$, and then decreases for $z > z_2$. This is the case for gamma-ray bursts (GRBs)~\cite{Wanderman:2009es}, the star formation rate (SFR)~\cite{Yuksel:2008cu}, and active galactic nuclei (AGNs)~\cite{Hasinger:2005sb}. Some tidal disruption event (TDE) and BL Lac models, on the other hand, allow for an overall flat or negative evolution of the emissivity~\cite{Hopkins:2006fq,Sun:2015bda,guepin2018a,Caccianiga:2001vu,Ajello:2013lka,AlvesBatista:2017shr}. We approximate and generalize the redshift evolution of these different source classes with the function given in Eq.~\ref{eq:sourceevolution0}. We do not take into account contributions from $z > 4.0$ as there the source evolution function for all source classes mentioned here is decreasing rapidly. Therefore, the contribution to the neutrino flux from this redshift range is expected to be negligible.

The sources are assumed to have an injection spectrum
\begin{equation}
	\frac{\text{d}N}{\text{d}E} \propto E^{-\alpha} \exp\left(  - \frac{E}{E_\text{max}} \right).
  \label{eq:spec0}
\end{equation}
We use the EBL model by Franceschini {\it et al.}~\cite{Franceschini:2008tp}. Nevertheless, for the energy range of interest ($E_{\nu} \gtrsim 1 \; \text{EeV}$ for neutrino energies), the CMB is the dominant photon field for neutrino production, thus implying that the choice of EBL model has a negligible effects, as shown in Ref.~\cite{alvesbatista2018b}. We perform the simulations in one dimension, i.e., neglecting magnetic fields. Magnetic-field effects might increase the expected cosmogenic neutrino flux by up to a factor of a few at $E_{\nu} = 1 \; \text{EeV}$~\cite{Wittkowski:2018giy}, depending on the assumed magnetic-field model and source distribution. Therefore, our predictions are rather on the conservative side.

For a fixed proton fraction  $f$ the only parameters that can be varied in our model are: $\alpha$, $E_\text{max}$ and $m$. We adopt the following ranges for them: $1.0 \leq \alpha \leq 3.0$, $19.6 \leq \log(E_\text{max} / \text{eV}) \leq 23.0$ and $-6.0 \leq m \leq 7.1$. This choice of parameter ranges encompasses spectral indices, maximum energies, and source evolutions found in many theoretical models for cosmic-ray sources. The maximum energy, however, is relatively high compared to the low $R_\text{max}$ scenarios found in recent phenomenological interpretations of the data~\cite{Aab:2016zth,Romero-Wolf:2017xqe,AlvesBatista:2018zui}. Nevertheless, because intrinsic properties of cosmic accelerators may vary significantly across members of a population of sources (while these phenomenological studies assume identical sources throughout the Universe), it is not unreasonable to expect that individual sources could have a higher $R_\text{max}$. To give an indication for how the results depend on the spectral index and maximum energy, and to show what happens for the most commonly used spectral indices, we additionally provide the outcomes for a more restrictive parameter range of $2.0 \leq \alpha \leq 3.0$ and $20.0 \leq \log(E_\text{max} / \text{eV}) \leq 23.0$. A more detailed investigation of the effects of each of the parameters on the flux of cosmogenic neutrinos can be found in Refs.~\cite{Kotera:2010yn,vanVliet:2016dyx,vanVliet:2017obm}. 

\begin{figure}[tbh]
  \includegraphics[width=0.95\columnwidth]{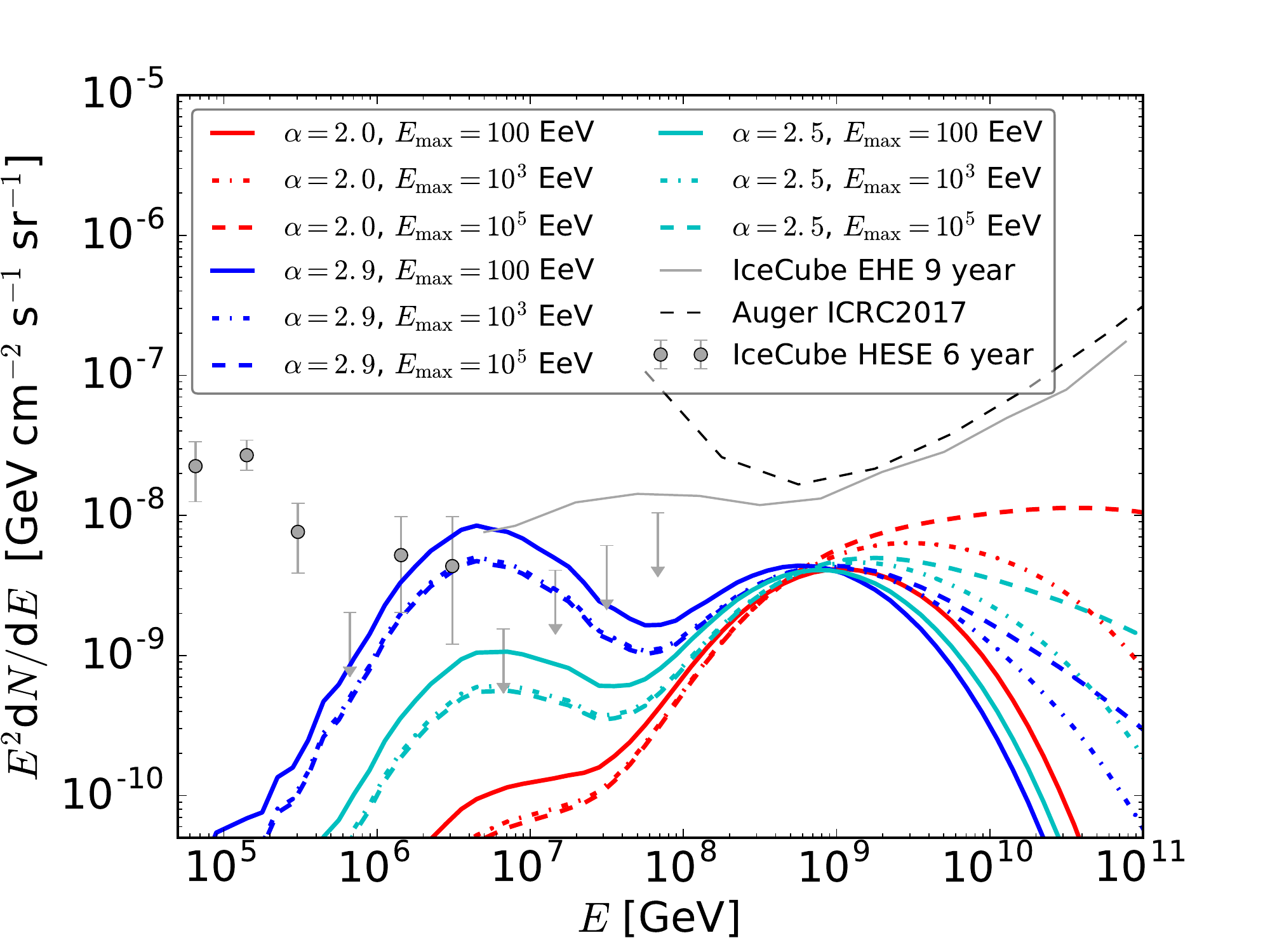}
  \caption{Simulated single-flavor cosmogenic neutrino ($\nu + \bar{\nu}$) spectra (assuming a $(\nu_e:\nu_\mu:\nu_\tau) = (1:1:1)$ flavor ratio) for pure-proton scenarios with $m=3.0$ and $f=1.0$. The corresponding cosmic-ray curves are normalised to the Auger spectrum~\cite{Fenu:2017hlc} at $E_0 = 10^{19.55} \; \text{eV}$. For reference, we also show the IceCube 6-yr HESE data~\cite{Kopper:2017zzm} and the Auger~\cite{Aab:2015kma,Zas:2017xdj} and IceCube~\cite{Aartsen:2018vtx} differential $90\%$ C.L. upper limits for single-flavor neutrinos and half-energy-decade fluxes.}
  \label{fig:spec}
\end{figure}

Note that for $17.5 \lesssim \log(E_{\nu} / \text{eV}) \lesssim 18.5$  the neutrino spectrum is roughly unaffected by the choice of $\alpha$ and $E_\text{max}$, provided that the latter is not too low. This is shown in Fig.~\ref{fig:spec} for the specific case of $f = 1.0$ and $m=3.0$, but similar behavior is seen for other values of $m$ and $f$. These neutrino spectra can straightforwardly be scaled down to get the results for smaller proton fractions, neglecting the subdominant contribution to the cosmogenic neutrino flux from heavier nuclei. So, if we focus on this energy range, the only two parameters that still have a significant effect on the expected cosmogenic neutrino flux are $m$ and $f$.

\begin{figure}[tbh]
  \centering
  \includegraphics[width=0.95\columnwidth]{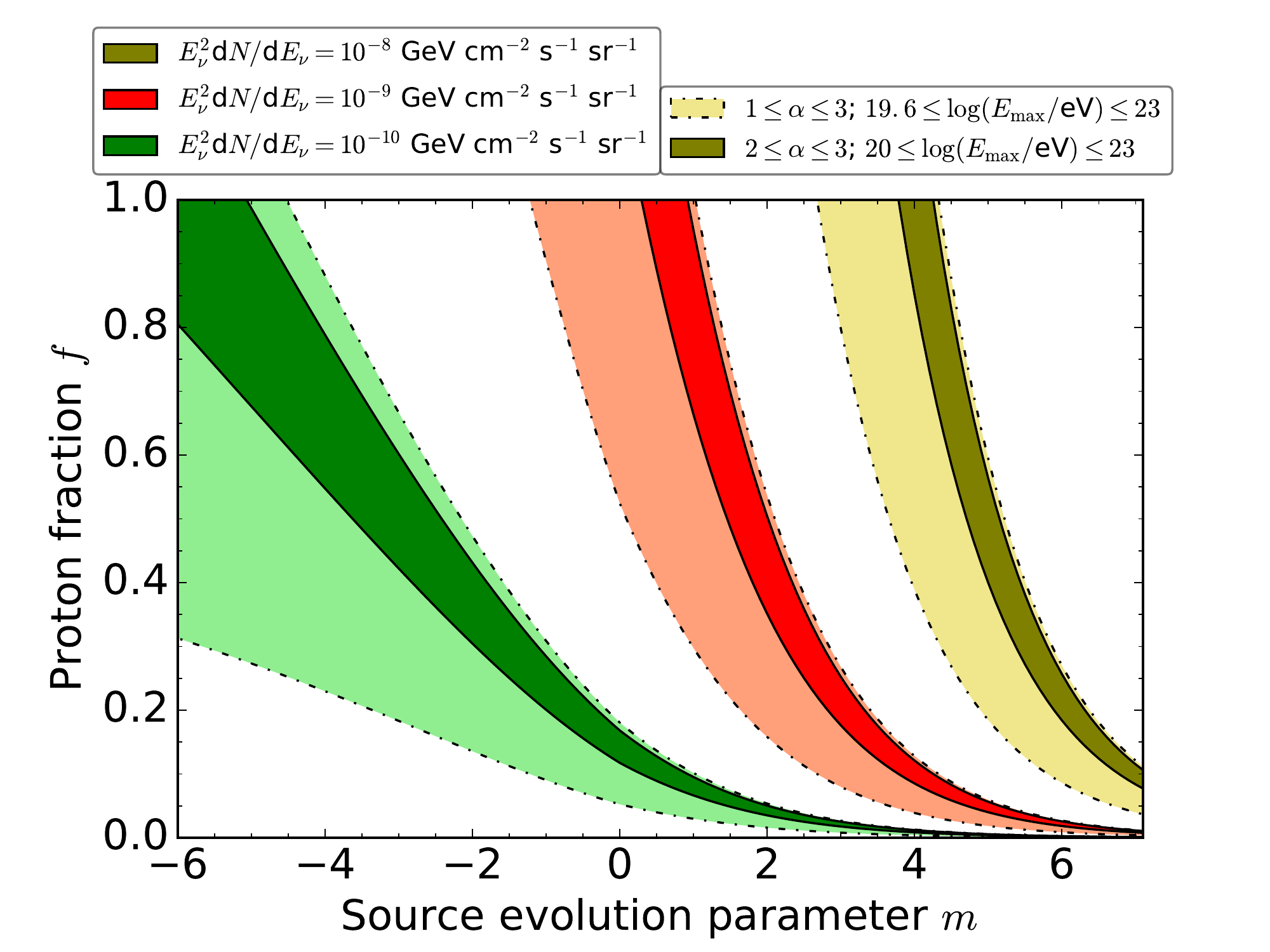}
  \caption{Observable fraction of protons $f$ at ultra-high energies as a function of the source evolution parameter, $m$. Three different single-flavor flux levels at a neutrino energy of $E_{\nu}=1$ EeV are shown, corresponding roughly to the current sensitivity of IceCube and Auger (yellow), and upper (red) and lower (green) ranges for the expected sensitivity of ARA, ARIANNA and GRAND200k.
  }
  \label{fig:pFrac}
\end{figure}

In Fig.~\ref{fig:pFrac} the proton fraction is plotted as a function of the source evolution parameter. Here each shaded area corresponds to a particular level of the cosmogenic neutrino flux at $E_{\nu}=1$ EeV and encloses all of the combinations of $m$ and $f$ that yield that flux level. The width of the shaded areas results from varying $\alpha$ and $E_\text{max}$ within the indicated ranges.

Auger and IceCube have set their current upper limits on the neutrino flux at $E_{\nu}=1$ EeV to $E_{\nu} ^2 \text{d}N/\text{d}E_{\nu} \simeq 10^{-8} \; \text{GeV} \, \text{cm}^{-2} \, \text{s}^{-1} \, \text{sr}^{-1}$, which corresponds to the yellow band in Fig.~\ref{fig:pFrac}. 
Therefore, it can be concluded that at the moment sources following a strong source evolution, $m \gtrsim 6.0$, that would produce a proton fraction of $f \gtrsim 0.27$ are already ruled out. 

Furthermore, future neutrino experiments will scan a significant region of the parameter space shown in Fig.~\ref{fig:pFrac}. ARA~\cite{ara2012a}, ARIANNA~\cite{arianna2015a} and GRAND200k~\cite{Alvarez-Muniz:2018bhp}
will nominally reach sensitivities of $E_\nu^2 \text{d}N/\text{d}E_\nu \sim 10^{-9} - 10^{-10} \; \text{GeV} \, \text{cm}^{-2} \, \text{s}^{-1} \, \text{sr}^{-1}$. A non-detection of cosmogenic neutrinos with a sensitivity of $\sim 10^{-9} \; \text{GeV} \, \text{cm}^{-2} \, \text{s}^{-1} \, \text{sr}^{-1}$ would constrain the proton fraction to $f \lesssim 0.20$ for $m \gtrsim 3.4$. A sensitivity of $\sim 10^{-10} \; \text{GeV} \, \text{cm}^{-2} \, \text{s}^{-1} \, \text{sr}^{-1}$ would give the same constraint on $f$ for $m \gtrsim -0.2$.
Additionally, this shows that, for realistic source evolutions and only small amounts of protons at the highest energies, these experiments have a strong potential for measuring a cosmogenic neutrino flux at $E_{\nu}\approx1$ EeV.

While the composition and source evolution are degenerate quantities~\cite{Moller:2018isk}, a prior on one of these two can be chosen to determine the other. One way to do this is by focusing on specific candidates for UHECR source classes. AGN, for example, can be subdivided in different classes with different redshift evolutions for $z < z_1$, according to their luminosities: Medium-Low Luminosity AGNs (MLL), Medium-High Luminosity AGNs (MHL) and High Luminosity AGNs (HL) (low luminosity AGNs are not expected to be able to accelerate cosmic rays up to ultra-high energies~\cite{Waxman:2004ww}). In table~\ref{tab:sourceconstraints} the constraints on $f$ are given for these and other possible source classes in case no neutrinos are detected at $E_{\nu}\approx1$ EeV for different flux levels. Only for High Synchrotron Peaked BL Lacs (HSP) will it be difficult for ARA, ARIANNA and GRAND to constrain the proton fraction (depending on the values of $\alpha$ and $E_\text{max}$).

\begin{table}[tbh]
\centering
\caption{Maximum proton fraction $f^{\text{max}}_n$ for different source classes, with their redshift evolution parametrized by $m$ (see Eq.~\ref{eq:sourceevolution0}), for non-detection at a flux level of $E_\nu^2 \text{d}N/\text{d}E_\nu = 10^{n} \; \text{GeV} \, \text{cm}^{-2} \, \text{s}^{-1} \, \text{sr}^{-1}$at $E_{\nu}\approx1$ EeV.}
  \begin{tabular}{l c c c c c c c}
  \hline
      & HL & MHL & MLL/SFR & GRB & BLLac & TDE & HSP  \\ 
   \hline
     $m$ & $7.1$ & $5.0$ & $3.4$ & $2.1$ & $0.0$ & $[0.0,-3.0]$ & $-6.0$    \\ 
     $f^{\text{max}}_{-8}$ & $0.11$ & $0.59$ & $1$ & $1$ & $1$ & $[1,1]$ & $1$  \\ 
     $f^{\text{max}}_{-9}$ & $0.01$ & $0.06$ & $0.20$ & $0.50$ & $1$ & $[1,1]$ & $1$  \\ 
     $f^{\text{max}}_{-10}$ & $0.00$ & $0.01$ & $0.02$ & $0.05$ & $0.18$ & $[0.18,0.67]$ & $1$  \\ 
  \hline
  \end{tabular}
  \label{tab:sourceconstraints}
\end{table}

While the reconstruction of neutrino showers does require some understanding of high-energy interactions with the atmosphere, the problem of uniquely identifying the composition of a cosmic ray is evaded by using neutrinos. Therefore, the method for determining the fraction of protons in UHECRs proposed here does not suffer from the large uncertainty in predicting $X_\text{max}$ from different hadronic interaction models. 

Additionally, this method can be used to determine the evolution of UHECR sources by combining the cosmogenic neutrino measurements with UHECR composition measurements. Auger already showed that, for $E \gtrsim 30 \; \text{EeV}$, $f \lesssim 0.20$, assuming Sibyll 2.1~\cite{Ahn:2009wx}, QGSJET II-04~\cite{qgsjetii} or EPOS-LHC~\cite{eposlhc} as hadronic interaction model and fitting a mixture of protons, helium nuclei, nitrogen nuclei, and iron nuclei~\cite{Aab:2014aea}. AugerPrime will significantly improve these results. With these measurements  and a detection of a cosmogenic neutrino flux at  $E_{\nu}\approx1$ EeV the evolution of UHECR sources can be established using Fig.~\ref{fig:pFrac}. As UHECR source candidates have a widely varying range of possible source evolutions this could lead to determining what the most likely source class for UHECRs is. 

It is important to stress that our results only hold for cosmogenic neutrinos, as opposed to neutrinos produced via photohadronic, photonuclear, or hadronuclear interactions of UHECRs with the surroundings of a source. This degeneracy has to be broken before any reliable constraint on the proton fraction is derived. It might be possible to do this by discerning the shape of the spectrum for $0.1 \lesssim E_{\nu}/\text{EeV} \lesssim 1$, which is typically harder than $E_{\nu}^{-2}$ for cosmogenic neutrinos when $f>0$ (see Fig.~\ref{fig:spec}). In addition, it might be possible to remove neutrinos originating from identified point sources from the cosmogenic neutrino flux. In this case a good angular resolution will be necessary if one wants to discern cosmogenic neutrinos from cosmic-ray protons originating in those point sources from neutrinos produced in interactions in the surroundings of the source, as the deflection of protons with $E_p \gtrsim 30 \; \text{EeV}$ by intergalactic magnetic fields could be small~\cite{Hackstein:2016pwa, AlvesBatista:2017vob} (depending on source distance and magnetic-field model).

In summary, we have presented a method to constrain the fraction of UHE protons arriving at Earth, based on cosmogenic neutrino fluxes. This method is robust in the sense that it does not directly rely on the inference of the composition of primary cosmic rays through the showers they induce in the atmosphere; instead, it relies on the identification of neutrino-induced air showers, whose signatures are more clear. The constraints that can be derived, however, do depend on assumptions regarding the redshift evolution of the source emissivity. Nevertheless, for most typical source evolutions, a proton fraction of $f \lesssim 0.20$ can definitely be constrained with future detectors such as ARA, ARIANNA and GRAND, provided that they reach their projected sensitivities. For strong source evolutions the current limits of IceCube and Auger already constrain the proton fraction to $f \lesssim 0.20$.

\vspace{2mm}

We would like to thank Anatoli Fedynitch for his comments on the paper and the participants of the 2017 SRitp workshop ``High-energy neutrino and cosmic-ray astrophysics - The way forward'' at the Weizmann Institute of Science, Israel for useful discussions.
AvV acknowledges financial support from the NWO Astroparticle Physics grant WARP and the European Research Council (ERC) under the European Union's Horizon 2020 research and innovation programme (Grant No. 646623).
RAB is supported by grant \#2017/12828-4, São Paulo Research Foundation (FAPESP).

\bibliography{references}

\end{document}